%% file: paper.tex
\documentclass[AMA,STIX1COL]{WileyNJD-v2}

\articletype{}%

\received{26 April 2016}
\revised{6 June 2016}
\accepted{6 June 2016}

\usepackage{booktabs} 
\usepackage{amsmath,amsthm,amssymb,mathtools}
\usepackage{enumitem}
\usepackage{algorithm}
\usepackage{algpseudocode}
\usepackage{listings}

\lstdefinelanguage{JavaScript}{
  keywords={jcond, jdata, jsync, function, for, size, jasync, var, typeof, new, true, false, catch, logger, char, struct, broadcaster, inflow, outflow, return, null, catch, switch, if, in, while, do, else, case, break},
  keywordstyle=\color{blue}\bfseries,
  ndkeywords={class, var, export, boolean, int, throw, double, implements, import, this},
  ndkeywordstyle=\color{purple}\bfseries,
  identifierstyle=\color{black},
  sensitive=false,
  comment=[l]{//},
  morecomment=[s]{/*}{*/},
  commentstyle=\color{orange}\ttfamily,
  stringstyle=\color{red}\ttfamily,
  morestring=[b]',
  morestring=[b]
}

\begin{document}

\title{A Language for Programming Edge Clouds for Next Generation IoT Applications}

\author{Muthucumaru Maheswaran}

\author{Robert Wenger}

\author{Richard Olaniyan}

\author{Salman Memon}

\author{Olamilekan Fadahunsi}

\author{Richboy Echomgbe}

\authormark{M. Maheswaran \textsc{et al}}

\abstract[Abstract]{
For effective use of edge computing in an IoT application, we need to partition the application into tasks and map them into the cloud, fog (edge server), device levels such that the resources at the different levels are optimally used to meet the overall quality of service requirements. In this paper, we consider four concerns about application-to-fog mapping: task placement at different levels, data filtering to limit network loading, fog fail-over, and data consistency, and reacting to hotspots at the edge. We describe a programming language and middleware we created for edge computing that addresses the above four concerns. The language has a distributed-node programming model that allows programs to be written for a collection of nodes organized into a cloud, fog, device hierarchy. The paper describes the major design elements of the language and explains the prototype implementation. The unique distributed-node programming model embodied in the language enables new edge-oriented programming patterns that are highly suitable for cognitive or data-intensive edge computing workloads. The paper presents result from an initial evaluation of the language prototype and also a distributed shell and a smart parking app that were developed using the programming language.
}


\maketitle

\input{intro}

\input{motivation}

\input{progmodel}

\input{jamlang}

\input{implementation}

\input{evaluation}

\input{patterns}

\input{applications}

\input{related}

\input{conclusions}

\bibliography{jamscript}

\end{document}

%% file: intro.tex
\section{Introduction}
\label{sec1}


The continuous popularity of cloud computing has created a large concentration of computing power at the core of the Internet. This has many well-documented~\cite{satyanarayanan2015edge, shi2016promise} problems for the Internet of Things applications that generate data at high rates and want to process them with short response times. Edge computing is a recent initiative to ease this concentration by building peripheral infrastructure that would place computing resources closer to the devices. Various studies have shown the potential of edge computing to reduce response times, lower bandwidth usage, enhance the energy efficiency, and so on. With the recent and escalating interest in smart cities and self-driving cars, a computing infrastructure at the edge that can be quickly leveraged by the devices (that is vehicles, sensors, and actuators) is sorely needed. For instance, self-driving cars generate lots of data using a variety of sensors attached to them. By sharing a portion of that data with edge resident servers, self-driving vehicles can enable powerful edge analytics that can be useful for public safety, augmented reality, and even self-driving cars~\cite{hu2015mobile}.

The importance of a shared computing platform between smart cities and self-driving cars is already recognized, and efforts are underway for creating {\em Transportation Mobility Clouds} \footnote{https://medium.com/cityoftomorrow/why-were-working-with-autonomic-to-create-a-platform-that-can-power-future-cities-96700c2824e6}. Creating TMCs with edge and cloud resources will bring us closer to realizing the ultimate pervasive computing vision~\cite{satyanarayanan2001pervasive} at city scales with fast moving objects like cars. To achieve such large-scale platforms combining clouds and edge, we need to solve many interesting problems including: performing a handover of devices (i.e., vehicles or mobiles) from one edge server to another as they move through space, dealing with edge server (fog) failures, and maintaining data consistency across the platform, and handling temporary network disconnections. This paper describes a programming language and middleware we designed and implemented for creating large-scale platform consisting of resources at the cloud, fog, and device levels.

Our language is heavily inspired by the ``software-defined networking'' (SDN) paradigm~\cite{mckeown2009software, kirkpatrick2013software} that allows the network operators to control the networks programmatically~\cite{bosshart2014p4}. SDN implements a ``controller-worker pattern'' where programs running in the controller determine how the workers (switches) should process the data flows. In SDN, the controller-worker inter-operation is realized using either a standardized protocol like OpenFlow \cite{mckeown2008openflow} (a bottom-up approach) or an embedded protocol definition language like P4~\cite{bosshart2014p4} (a top-down approach). Our language uses a controller-worker model like SDN. However, instead of relying on a standardized protocol for inter-operation, the controller and worker parts are synthesized from a single program by the compiler leveraging the elements of the language design. In SDN, determining the number of controllers and their locations are important problems that can impact performance and fault tolerance~\cite{lantz2010network}. A large-scale platform that admits mobile elements can be even more sensitive to controller locations. To address this problem, we use a hierarchy of controllers. Therefore, each program can be considered a tree with workers at the leaves and controllers elsewhere. A problem is solved by a horizontal composition of trees with each implementing a solution for a sub-problem.

To illustrate the tree model for programs, let's consider a smart parking application (an example we implemented in the language and described in Section~\ref{samples}). The objective of parking. We can decompose this problem into three sub-problems: sensing for free spots, allocating spots for requests, and routing requests from the cars to the spot allocators.
We develop one application for each of the sub-problems. So a tree will be responsible for aggregating the parking spot availability information. The leaves of the sensing tree will run in
the sensors located at each parking spot, which notify the status of the spot. The root of the sensing tree will run in the cloud and will hold the global status information which may be in aggregate
form.

Similarly, the allocation application is organized as a tree that runs in the cloud, fogs, and devices. The devices could be street-level management stations. The third application (represented as yet another tree) could be the request broker that routes or aggregate the requests from the cars that seek parking. With the three trees mapped onto the same cloud, fog, and device nodes, the applications can talk with each other even while disconnections exist in the whole resource pool (i.e., some fogs along with their devices are disconnected, or some devices are disconnected from their associated fogs). The language provides primitives for creating inter-application data flows that can be used for the allocator to the spot sensor in the smart parking problem.

Although the tree-based controller-worker model we have for each application can have many levels (to accommodate multiple fog levels~\cite{tang2015hierarchical}), we consider a three-level tree corresponding to cloud, fog, and device (mobile) levels like several recent studies~\cite{hong2013mobile, luan2015fog, stojmenovic2014fog}. Because a device has many possible fogs it could associate with to form the tree,
there are interesting fog attachment (initial) and reattachment (in the event of a failure or mobility) problems. In the simplest case, these problems are
solved by merely considering fog and device proximity. For fault tolerance, we need to replicate the immutable data objects held in the fogs, which creates data movement costs.

Our work makes the following contributions towards the development of
large-scale edge-centric computing platform.
\begin{itemize}
    \item
    We developed a distributed-node programming model for the cloud, fog, and
    device hierarchy that can create horizontally composable
    programs. The programming model is based on distributed memory; hence, data flow mechanisms are provided to
    realize intra- and inter-application data exchanges
    irrespective of the physical node mappings.

    \item
    We designed and implemented a polyglot language that implements the distributed-node
    programming model. The language was used to implement several example applications, including a smart parking system and a distributed shell that runs on the cloud, fogs, and devices.

    \item
    We formulated several edge-centric programming patterns that leverage the unique features of the language. The programming patterns provide guidelines for the programmer to solve important problems like task placement and data feeding rate from sensors.

    \item
    We carried out several performance evaluation experiments on the language runtime and on the applications developed using the language. A Docker container-based emulator we developed was used to create various cloud, fog, and device configurations for the experiments.
\end{itemize}

Section~\ref{motivate} provides the motivations and challenges for this research work.
The core programming model of the language is provided in Section~\ref{pmodel}.
Section~\ref{ldesign} describes the design and implementation of the language.
An optimization model for locating the fogs in the best server locations is
given in Section~\ref{optmodel}.
Experimental results are presented in Section~\ref{results}.
Several edge-oriented programming patterns that are enabled by our programming
language is given in Section~\ref{patterns}. Two sample applications are
discussed in Section~\ref{samples}. Related work is described in Section~\ref{related}.

%% file: motivation.tex
\section{Motivating Scenarios and Challenges}
\label{motivate}

In this section, we describe several scenarios that motivate the different design features of our language. Important challenges that need to be tackled while implementing these features are also discussed.

\subsection{System levels \& Function Placement}
A triple-leveled system with cloud, fog, device levels is widely~\cite{garcia2015edge} accepted as the edge computing reference architecture. In this architecture, cloud level is expected to handle functions that need to execute with global perspective while the fogs handle functions that need low response time and can run in local scope. The device level would run functions that are very critical but does not need remote data.
In certain applications, the programmer may want to place the function statically at a given level. For instance, a directory look-up for people or devices should run at the global scope. However, such a function would fail to execute if the device is partitioned from the cloud.
A pothole information collector function, on the other hand, would want to run in the local context (at the fog level) because information from a remote context is not very useful. The above examples illustrate static function to level assignment. In other cases, like a smart parking application, the parking spot search function needs to run at different levels depending on availability of parking spots and price the requester is willing to pay.

\subsection{Adaptive Data Probing}

The high-rate data updates from IoT is one of the primary motivators for edge computing~\cite{shi2016promise}.
With edge computing, the high-speed data updates are stored and processed at the edge; thus, alleviating
the network bottlenecks the data updates could create if they are sent to the clouds. Because edge resources have limited capacities and many applications from different users would want to use them, it is important to use
the edge resource very efficiently~\cite{liang2017mobile}.
One way to reduce the load on the edge itself is to throttle down the
data update rates to reduce the storage and processing requirements. The actual data update rate an IoT should use is highly application dependent. However, the amount of throttling we need to employ is dependent on the resource scarcity because the applications would like to go at the highest possible fidelity if there are plenty of resources.

For example, consider a heating, ventilation and air condition (HVAC) system which receives temperature, humidity and other measurements from a large number of sensors. To reduce the volume of data and processing overheads, the update rates could be set low. Suppose the system is reporting abnormal fluctuations in temperature or other parameters, we would want to increase the update rates to obtain large data traces and analyze them to detect possible root causes. Similarly, we could adjust sensing rates across different seemingly sensing zones that could shed light on each other in a root cause analysis. For instance, increasing traffic collisions at a road segment might need increased sensing at the road segments that bring in and lead away traffic from the given segment. The key requirements of adaptive data probing are adjusting the update rates of a sensor and selecting the sensors to adjust according to the application requirements and available edge resource capacities.

\subsection{Fog Fault Tolerance}

Fog resources that are placed at the edge and experiencing harsher conditions are bound to fail more than cloud resources that operate under highly controlled environments. Fog installations consist of small clusters of servers that are physically separated from each other, where the primary placement objective is to remain
close to the eventual devices (mobiles or fixed) served by them. Because the different servers in a fog cluster experience the same harsh conditions, we can expect them to have a dominant common mode failure. That means to backup a fog server we need to select another fog from another cluster resulting in a relatively large latency between them. Therefore, typical cloud fault tolerant solutions (e.g., active state machine replication~\cite{bessani2014state}) are not applicable to fogs.

Another fault-tolerant technique widely used in batch and stream processing systems is check-pointing and restarting~\cite{bala2012fault}. Because edge computing could be used for real-time applications such as operation control of IoT (e.g., drones), restarting from past checkpoints may not be appropriate; instead, restarts need to be performed from currently valid operating points.

Edge computing primarily positioned as a performance enhancer or data/service localizer for the already dominant cloud computing paradigm. Therefore, fault tolerance should be added to fog resources while maintaining the performance and locality benefits. When a device is attached to a fog, it could upload significant amount of data to the server and the fog could apply processing operations on the data. So when the fog fails, one way to recover is to find another fog that could provide the lowest cost restoration of device's operating point to the previous one (e.g., lineage graphs~\cite{zaharia2012resilient}).

\subsection{Edge Load balancing \& Unbalancing}

Load balancing is an age-old problem~\cite{amis2000load} in server clusters; however, edge computing brings a new flavor because we need a low-latency scheduler~\cite{ousterhout2013sparrow} that is locality-aware and fault tolerant. Furthermore, task scheduling itself needs to be decentralized so that it could proceed without {\em completely} relying on a central scheduler for optimal decisions.

For example, with a smart parking application, a car requesting parking needs to be connected to the fog that has most current information regarding the local parking spaces. At the same time, we need a fog that is not overly loaded because otherwise the car would be wasting precious time in the fog and lose available spots to late arriving cars. So the predicted loading of the fogs could be used to deploy multiple servers for a particular request zone.
Multiple applications with different quality of service (QoS) requirements could be hosted by a fog resource at the same time. The load balancing requirements of the co-hosted applications could be quite diverse. Therefore, application-level coordination of load balancing is necessary.

In emergency situations, devices in a certain zone or devices of a certain type could become more important (e.g., in natural disasters or terrorism-related situations, emergency services will have escalated privileges for resource access). The fog resource allocation for the devices with urgent applications should get precedence over other
devices' requests. That is, we need controlled unbalancing schemes that would provide differential QoS to cater for emerging situations at the edge.

%% file: progmodel.tex
\section{Programming Model}
\label{pmodel}

\subsection{Controller-Worker Programming Model}

As a straw man considers a controller-worker model, where a single controller is managing many workers. A program written for such a model contains two types of functions: {\em local} and {\em remote}. The local functions run either in the controller or worker. They consume execution time and need to be scheduled according to the synchronization requirements of the remote functions. The remote functions can be either controller-to-worker (Ctrl2Wrk) functions or worker-to-controller (Wrk2Ctrl) functions. The Ctrl2Wrk functions run on all workers underneath the controller. Similarly, the Wrk2Ctrl functions are invoked by the workers and run on the controllers. The Wrk2Ctrl and Ctrl2Wrk functions can be either {\em synchronous} or {\em asynchronous}. With synchronous functions, the caller will block until the remote function returns the results. The asynchronous functions, on the other hand, do not block; the caller is free to execute the next statement after the remote execution launches.

The controller and workers are independent nodes with independent address spaces. One way of exchanging data is through parameters for remote function calls. In many cases, IoT have time series data and update rates could be quite high. Remote function call parameters is not the best way of exchanging this type of data between the workers and controller.

\subsection{Tree of Controllers and Varied Deployments}

\begin{figure}[!ht]
\centering{\includegraphics[width=3.0in]{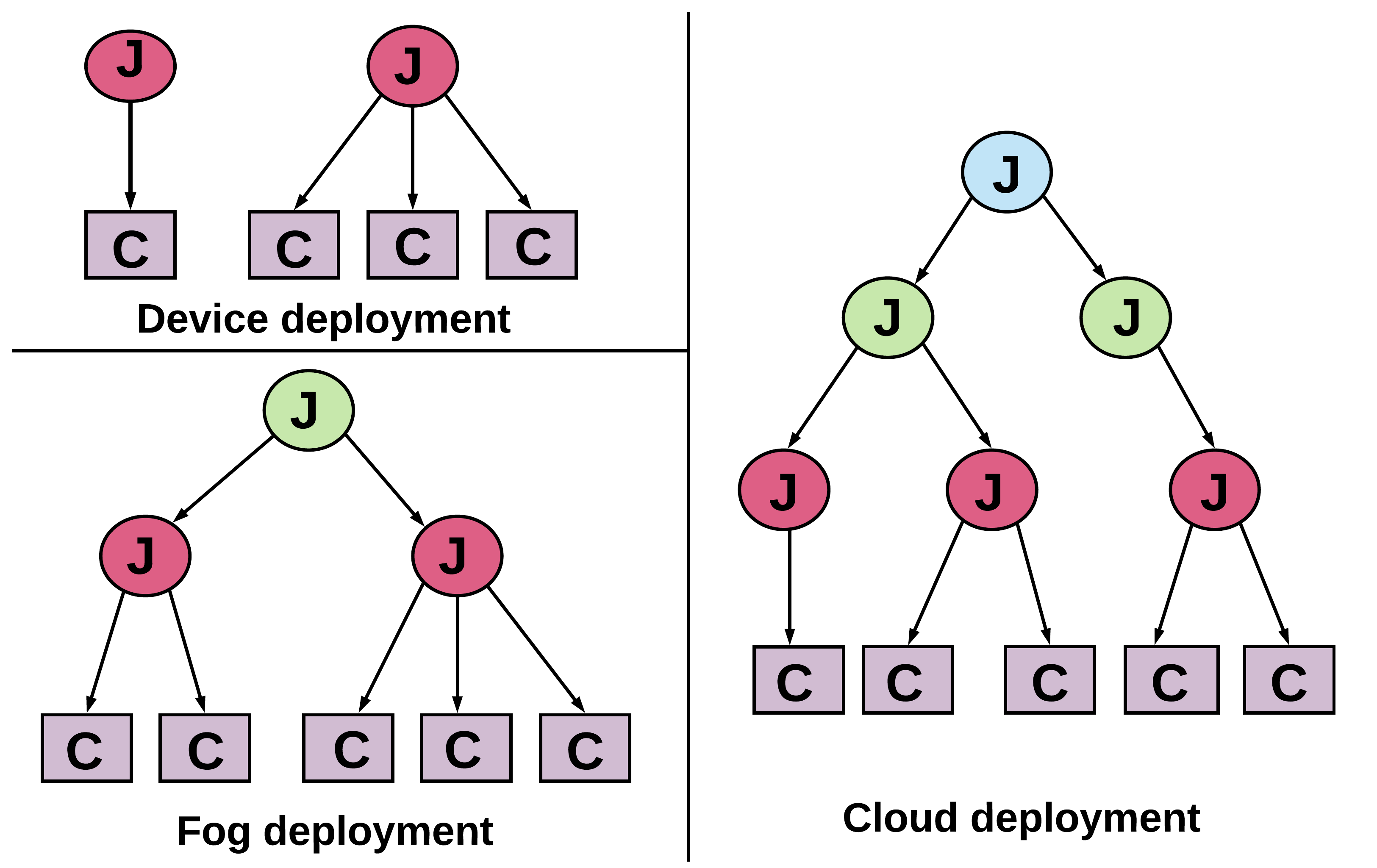}}
\caption{Different tree deployment models.}
\label{models}
\end{figure}

The single controller model could work with cloud computing. However, with edge computing, the single
controller would make either the cloud or the fog not very useful. Therefore, we extend the controller to a tree of controllers. With this model, the root of the controller tree could be in cloud and leaves could connect to the workers. With a controller tree, the edge can host a controller at a subordinate level. Even the devices can host a subordinate controller (of course below the edge-level subordinate controller) to provide deeper autonomy. Figure~\ref{models} shows different ways of deploying tree of controllers and workers, controllers are denoted with J and workers denoted with C. As shown, a program developed for the controller-worker model could be deployed in a single device for stand-alone execution. It could be deployed across a collection of devices under a single fog with the J (controller) running in the fog and another J (local controller) in the devices along with the workers. The most general execution is the one with nodes at all three levels.

With a tree of controllers, it becomes possible for a controller to invoke remote function on another controller. The programming model restricts the remote Controller-to-Controller (Ctrl2Ctrl) functions to a sub-machine that is rooted at the calling controller. That is a controller cannot make a remote call on another controller above its level.

\subsection{Data Flows in Controller-Worker Trees}

As discussed earlier, we need an efficient data exchange mechanism between the controller and workers. Particularly with IoT, the update rates from workers to controllers could be quite high. The time series data that is predominantly interchanged can have gaps~\cite{fekade2017probabilistic} in them.
We introduce two constructs to handle high-rate time-series data: loggers and broadcasters. The
logger can be considered as a pipe that takes data written by a worker to its controller and then to the super controller and so on. We could also have the first hop controller processing the data to that it does not propagate further in the form it was written by the worker. Similarly, the broadcasters provide data flow downwards from the controllers to sub controllers or workers. The loggers and broadcasters do not offer any reliability. There could be reliable or learning (that repair the data streams or predict the stream values ahead of time) implementations of these constructs.

\subsection{Inter-Application Data Exchanges}

Complex applications can benefit from multiple trees that are working together by interchanging data and results. It is a pipe construct like the logger for a controller to push data to another controller. Unlike logger or broadcaster which interchange data with controllers at different levels, this construct works only with controllers at the same level and between two different applications. That is, two controllers (for example, the controllers at the fog or device levels) of the same application cannot be part of this interchange.

%% file: jamlang.tex
\section{Language Design and Implementation}
\label{ldesign}

\subsection{Overview}
The language described here implements the programming model explained in Section~\ref{pmodel}. It is a polyglot language; it brings together C and JavaScript with some additional constructs needed to implement the programming model. The C side implements the {\em worker} side of the programming model and JavaScript side implements the {\em controller} side. This means the C side runs at the lowest level of the system. For instance, embedded devices (e.g., sensors and actuators) will be running the C component. This allows us to target memory constrained devices (planned for the future). The JavaScript running in a NodeJS runtime is responsible for implementing the controllers. The NodeJS based controller can run from low power devices to all the way to cloud servers.

The C side of our language is single threaded to match the JavaScript side. In the controller, if there is a workload that needs intensive processing, the pipes-and-filters pattern discussed in Section~\ref{patterns} should be used instead of holding the main thread.

\subsection{Major Elements of the Language}

\subsubsection{Remote Functions}
Any function either on the C side or JavaScript (J) side could be designated as a remote function by prepending the keyword {\tt jsync} or the keyword {\tt jasync}.
If the keyword {\tt jsync} is used, the remote function is a {\em synchronous} function.
That is the calling node blocks until the function returns a value. A function prepended with
{\tt jasync}, on the other hand, is {\em asynchronous} and is non-blocking. Also, there is no
return value from an asynchronous function. However, a callback can be passed into such a function at invocation for future notification from the function.

\subsubsection{Shared Persistent Variables}

Loggers provide the ability to store data as a time series. Loggers can be specified at the device, fog or cloud level. The data will be stored at the level specified, and can only be accessed from that level.
There are two types of loggers: cloud logger and fog logger.
\begin{lstlisting}[language=JavaScript]
jdata {
    	double x as logger(cloud|fog);
}
\end{lstlisting}

Writing to the logger from the C side is done by simply assigning to a declared logger variable as if it were any other variable that had already been declared. It can be used in any function that contains C code:
\begin{lstlisting}[language=C]
int testFunction() {
	x = 32;
}
\end{lstlisting}

Broadcasters provide a mechanism for pushing data from the cloud or fog to the devices. Unlike a logger which contains a time series, a broadcaster contains at most one value available to the program at any given time.
A broadcaster is defined with the {\tt broadcaster} keyword:
\begin{lstlisting}[language=JavaScript]
jdata {
	double x as broadcaster;
}
\end{lstlisting}
Broadcaster data originates on fog or cloud nodes. Conditions are used to control where the data originates. For example, the execution of an activity that writes a value to a broadcaster may be restricted only to fog nodes (vs device and cloud nodes).

Broadcaster data originating on a fog node:
\begin{lstlisting}[language=JavaScript]
jdata {
  	double x as broadcaster;
}
jcond {
  	fogonly: sys.type == "fog";
}
jasync {fogonly} function fname() {
  	x = fname1();
}
\end{lstlisting}

Broadcaster data originating on a cloud node:

\begin{lstlisting}[language=JavaScript]
jdata {
  	double x as broadcaster;
}
jcond {
  	cloudonly: sys.type == "cloud";
}
jasync {cloudonly} function fname() {
  	x = fname1();
}
\end{lstlisting}

When the broadcaster data is consumed, the latest value is always used. A function referencing a broadcaster does not execute until the node on which the program runs has received the latest value of the broadcaster. A condition referencing a broadcaster is always evaluated using the latest broadcaster value. In order to ensure producer-consumer synchronization, the runtime performs versioning of the broadcaster values.
Example:

\begin{lstlisting}[language=JavaScript]
jdata {
  	double pe as broadcaster;
}
jcond {
  	pickpe: pe < sys.rank;
 	 // evaluated using the latest broadcaster value
}
jasync {pickpe} function fname() {
  	double y = pe; // assigns the latest broadcaster value
}
\end{lstlisting}

\subsubsection{Conditions on Functions}

Conditions can be used to restrict the execution of an activity to certain nodes. For example, the execution of an activity may be restricted only to device nodes (vs fog and cloud nodes).

If a logger appears in a {\tt jcond} definition, then its latest value is used when the condition is evaluated. A logger is a collection of data streams each coming from a different device. Each of these data streams has a latest value. The latest among those latest values is used as the value for the logger when the condition is evaluated.
For example, consider a thermostat that is measuring the temperature in a room. The measurements are written to a logger temp. A function that executes periodically only on the devices turns on the heater when the temperature is too low.

\begin{lstlisting}[language=JavaScript]
// logger temp is in the global namespace
jdata {
	double temp as logger(fog);
}
// condition lowtemp is in the global namespace
jcond {
	lowtemp: sys.type == "dev" && temp < 18.5;
}
jasync {lowtemp} function fname() {
  // turns on the heater in the room
}
\end{lstlisting}










\subsection{Implementation of the Language}

To implement the language, we created a custom compiler that compiles the C and
JavaScript code with added constructs. The compiler outputs a custom executable
file that packages all the elements that are required to execute a program. A
loader instantiates the custom executable on MacOS, Linux, or Raspberry Pi. The
loader is also modified to work with docker containers so that large-scale
experiments can be carried out on a docker container-based emulator for
cloud-fog-device system. Our custom compiler is written using
Ohm~\cite{dubroy2017incremental}, a parser generator created by Alex Warth. The
middleware uses MQTT~\cite{hunkeler2008mqtt} and Redis (an in-memory data
store)~\cite{carlson2013redis}.

\begin{figure}[!ht]
\centering
\centering{\includegraphics[width=3.0in]{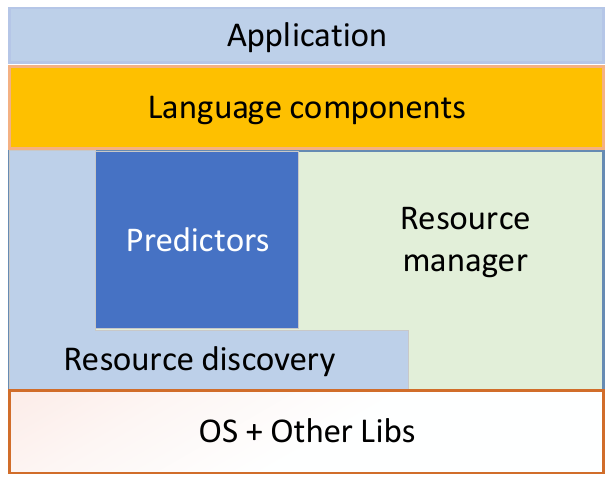}}
\caption{Key components of the language runtime stack.}
\label{lstack}
\end{figure}

Figure~\ref{lstack} shows the key components of the language runtime. One of the
foundational components is {\em resource discovery} (RD) that allows a program to discover
all elements that would already be running in the network and are potential
candidates to connect to. For instance, when a program is run in a device it
should see all the fogs that are already running the same program. The device
runtime selects the {\em best} fog among the available choices and connects to
it.

Primarily, the language offers support for three operations: remote functions
(i.e., RPCs), data management from  logging and broadcasting, and conditional
execution of remote functions. The remote functions are built on top  of the
publish-subscribe service provided by MQTT. One of the key additions is the time
aligning task scheduling that is required  to implement the synchronous Ctrl2Wrk
functions. The logging and broadcasting take place through the Redis in-memory
datastores. Redis provides very large persistent memory buffers where data could
be written into. When a worker writes a fog logger, it would be staged through
the intermediate Redis server at the device. Because the data is logged
initially to the Redis at the device and then pushed from there  to the Redis at
the fog, logging can tolerate disconnections. Also, we can optimize the
transfers to minimize the number of bytes transferred from the device to the
fog.

One of the important features we have in the language is the integration of
machine learning into the language runtime. Using the RD component runtime
determines the fogs that are available at different space-time points. We feed
this data into a {\em neural network} (NN) based predictor and train it so that
the NN-based predictor can tell the fogs a device would use in its journey ahead
of the device reaching a given point. We can use this feature to speculatively
change the association to the future work so that the handover  latency can be
minimized. Also, the any potential disconnections like a vehicle passing through
a tunnel could be detected ahead of time and the runtime can advance or delay
the launching of the remote tasks according to their criticality.

At the core the runtime is responsible for launching the remote function calls
and  scheduling the data uploads and downloads. In the simplest case, the
runtime can  immediately launch the requests and data transfers and attempt
retries in the event of failures. However, the system needs to more
sophisticated to take advantage of the many fog choices and the  changing
situation with regard to fog proximity. To handle this complexity, we have a
{\em resource management} (RM) component in the runtime. The RM expects the RD
to discover a pool of resources and the NN-based predictors  to forecast the
expected performance.

%% file: implementation.tex
\section{Optimizing Fog-Device Allocations}
\label{optmodel}



\begin{figure}[!ht]
\centering{\includegraphics[width=3.3in]{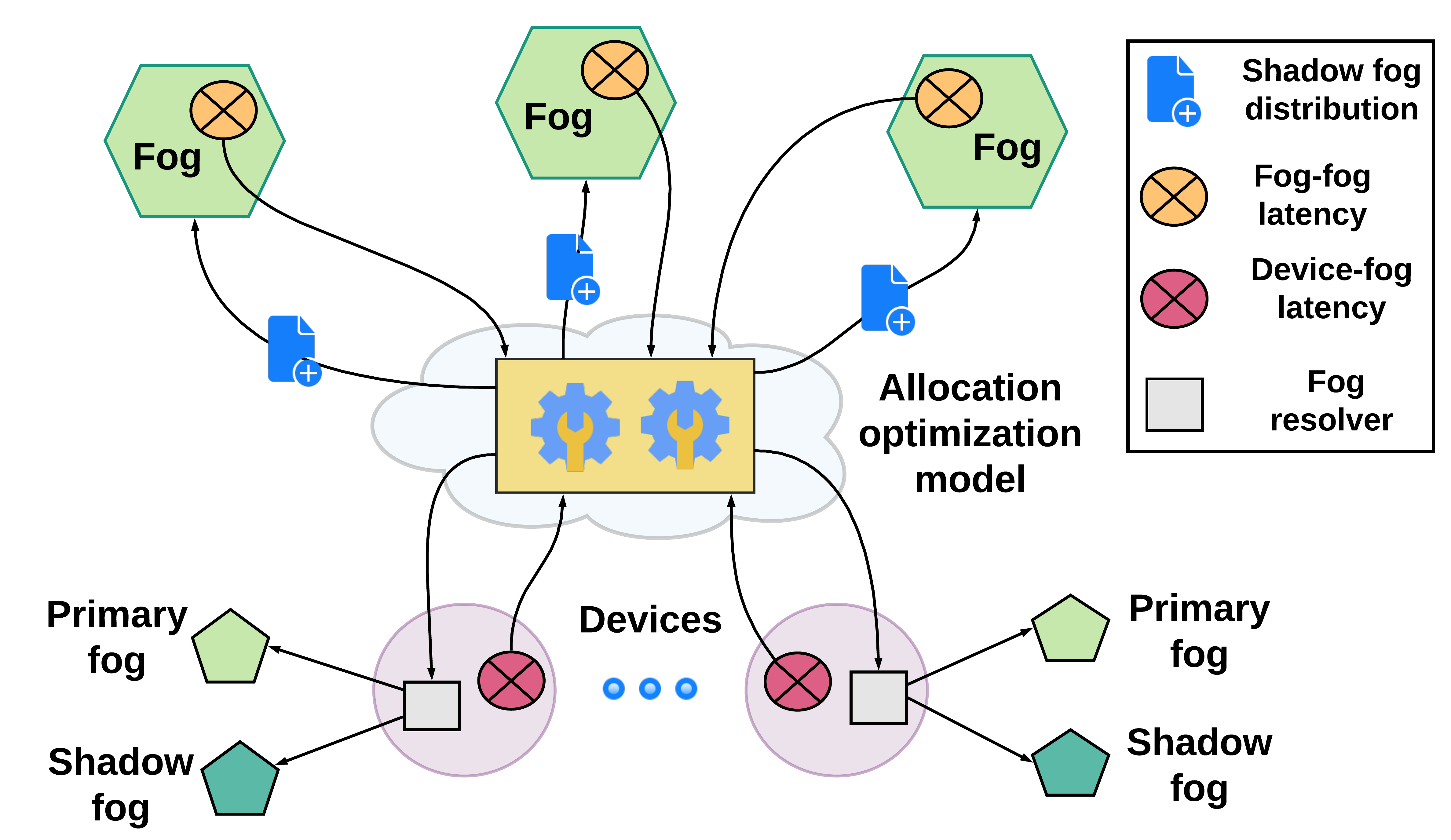}}
\caption{Allocation optimization model.}
\label{allocator}
\end{figure}

To understand the fog-to-device allocation problem, let's start with a configuration where all the devices are connected to the cloud. The devices are injecting data into the cloud via the logger constructs (defined in the program) and expect the cloud resident program (i.e., controller program) to operate on that data to create further data or actions that will steer the computations at the devices. We can model this situation by representing loggers by vertices of a graph where the edges represent the relations that can exist among the devices with regard to processing the data in a given application scenario~\cite{simmhan2015}. Now, let's say we want to introduce fogs into the system.
We can use the above graph to determine the best fog-to-device allocation. There are two considerations:
\begin{enumerate}
    \item We want all devices that are inter-related with regard to data processing (i.e., their logger data) in a given fog.
    \item We want to account for fog failures by maintaining replicas with certain level of consistency and put the replicas in nearby fogs.
\end{enumerate}
In this section, we formulate and solve an optimization problem that does just that.
The allocation optimizer process illustrated in Figure~\ref{allocator} provides the
design to implement this functionality into the language runtime.
The allocation optimizer essentially partitions a graph where the vertices represent the data objects (loggers) created by the devices and the edges represent the relationships among them. In addition, this graph holds shadow vertices that are additions created to represent replicas of the data objects created by the devices. As the name shadow implies, we don't expect the replica data object to be always current because maintaining fully consistent replicas is costly and could negate the performance advantage we are seeking from fog computing. To reduce the computation burden associated with maintaining replicas, we could even maintain an aggregate replica (i.e., the shadow vertex could be representing aggregate information about the actual data objects).

The optimizer runs in the cloud where it receives the device and fogs locations. It uses the device-to-fog latency and inter-fog latency to map the devices to fog while satisfying the capacity constraint of the fogs. It also gives the placement of the shadow vertices to the fogs.
A fog resolver that is embedded in the device receives the fog that the device should connect to called the primary fog and the corresponding shadow fogs. A shadow fog is an alternate fog to the primary fog that the shadow vertex is resident in.

The fog resolver routes the device request to the appropriate fog using a request routing algorithm. The algorithm primarily routes the request to the primary fog and in the event of the primary fog failure which is detected by the algorithm, it routes the request to the shadow fogs.




Mathematically, the allocation model is presented below. Let $D$ be the set of Devices and let $F$ be the set of Fogs.
Let $x_{ij}$ be the decision variable such that

\begin{equation*}
    x_{ij} = \begin{cases}
               1    & \text{if  device $i$ is mapped to fog $j$}\\
               0    & \text{otherwise}
           \end{cases}
\end{equation*}
and let $y_{jk}$ be the decision variable such that

\begin{equation*}
    y_{jk} = \begin{cases}
               1    & \text{if  shadow vertex of devices in fog $j$ is mapped to fog $k$}\\
               0    & \text{otherwise}
           \end{cases}
\end{equation*}

The latency cost between device i and fog j is given as $c_{ij}$ and the update cost for shadow vertices between fog j and fog k is given as $u_{jk}$.

Let $f_j$ be the fixed cost of using a fog service, $cap_{j}$ be the capacity of a fog j and the availability of a fog $a_j$ be defined as follows:
\begin{equation*}
    a_{j} = \begin{cases}
               1    & \text{if fog $j$ is available}\\
               0    & \text{otherwise}
           \end{cases}
\end{equation*}

The model is formulated as follows:

\begin{equation}
\begin{aligned}
& {\text{minimize}}
& &   z = \sum_{j \in {\cal F}} f_{j}a_{j} + \sum_{j \in {\cal F}}\sum_{k \in {\cal F}}u_{jk}y_{jk} + \sum_{i \in {\cal D}}\sum_{j \in {\cal F}}c_{ij}x_{ij}\\
& \text{subject to} & & \sum_{j \in {\cal F}} x_{ij} = 1, \hspace{101pt} \forall i \!\in \!{\cal D} \\
&  & & \sum_{j \in {\cal F}} y_{jk} = 2, \hspace{100pt} \forall k \!\in \!{\cal F}, j\neq k \\
& & &  \sum_{i \in {\cal D}} x_{ij} \leq cap_{j}* a_{j}, \hspace{74pt} \forall j \in {\cal F}\\
& & & x_{ij}, y_{ij}, a_j \in \{0,1\}
\end{aligned}
\end{equation}

The objective is to minimize the total costs involved while mapping the devices to fogs and ranking the placement of shadow vertices in fogs for each fog. In the objective function, the first term shows the cost of using an available fog, the second term the shows the cost of placing a shadow vertex in a fog and the third term shows the cost of mapping a device to a fog.  The first constraint states that each device must be assigned to exactly one fog and the second constraint states that a shadow vertex should be placed in two fogs such that the initial fog is not chosen. The third constraint states that for each fog, the capacity of the fog should be respected.
The model is solved using the IBM Cplex optimization solver.


A sample data set and solution of the allocation model is given below. Given 8 devices and 5 fogs, the device-to-fog latency and inter-fog latency is given in tables \ref{devfog} and \ref{fogfog} respectively.
The fogs have a fixed cost of service given as \$400 each, and we assume each fog to have a capacity of 3 however, the above parameters can be varied for each fog.

\begin{figure}[!ht]
\centering{\includegraphics[width=3.3in]{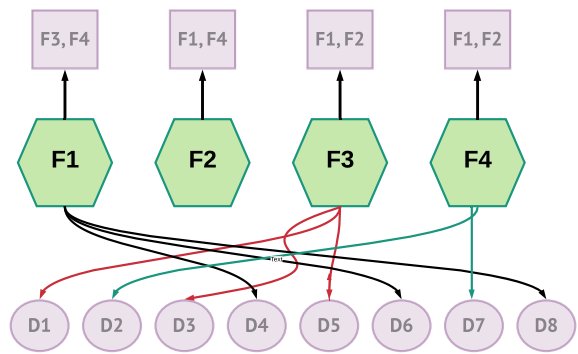}}
\caption{Optimization solution for sample data-set.}
\label{soln}
\end{figure}

The solution shows that three fogs should be employed and the distribution of devices to fogs is given in Figure \ref{soln}. Devices 1, 3 and 5 are mapped to Fog 3, while Fog 1 is mapped to devices 4, 6 and 8. Devices 2 and 7 are given to Fog 4. The shadow vertex for Fog 1 is placed in Fogs 3 and 4 while the shadow vertex for Fog 4 is placed in Fogs 1 and 2. Fog 2 and Fog 5 (not shown in the figure) are not available to devices.

\begin{table}
\caption{Device-to-Fog latency}
\centering
\label{devfog}
\begin{tabular}{|l|l|l|l|l|l|}
\hline
\multicolumn{1}{|c|}{} & \multicolumn{1}{c|}{Fog 1} & \multicolumn{1}{c|}{Fog 2} & \multicolumn{1}{c|}{Fog 3} & \multicolumn{1}{c|}{Fog 4} & Fog 5 \\ \hline
Device 1 & 24 & 74 & 31 & 51 & 84 \\ \hline
Device 2 & 57 & 54 & 86 & 61 & 68 \\ \hline
Device 3 & 57 & 67 & 29 & 91 & 71 \\ \hline
Device 4 & 54 & 54 & 65 & 82 & 94 \\ \hline
Device 5 & 98 & 81 & 16 & 61 & 27 \\ \hline
Device 6 & 13 & 92 & 34 & 94 & 87 \\ \hline
Device 7 & 54 & 72 & 41 & 12 & 78 \\ \hline
Device 8 & 54 & 64 & 65 & 89 & 89 \\ \hline
\end{tabular}
\end{table}

\begin{table}
\caption{Inter-Fog latency}
\centering
\label{fogfog}
\begin{tabular}{|l|l|l|l|l|l|}
\hline
\multicolumn{1}{|c|}{} & \multicolumn{1}{c|}{Fog 1} & \multicolumn{1}{c|}{Fog 2} & \multicolumn{1}{c|}{Fog 3} & \multicolumn{1}{c|}{Fog 4} & Fog 5 \\ \hline
Fog 1 & 1000 & 74 & 31 & 51 & 84 \\ \hline
Fog 2 & 57 & 1000 & 86 & 61 & 68 \\ \hline
Fog 3 & 57 & 67 & 1000 & 91 & 71 \\ \hline
Fog 4 & 54 & 54 & 65 & 1000 & 94 \\ \hline
Fog 5 & 98 & 81 & 16 & 61 & 1000 \\ \hline
\end{tabular}
\end{table}



%% file: evaluation.tex
\section{Experimental Results and Discussion}
\label{results}

We performed several experiments to evaluate the performance of the language and runtime as well as demonstrate some of its features. The experiments were conducted using a docker container-based emulator we have developed to run cloud-fog-device topologies with given inter-node latencies to mimic realistic scenarios on a machine running Ubuntu 16.04 with a Intel Xeon CPU with 16 cores and 128GB of RAM. For these experiments, in the emulator, the device to fog latencies are modelled by a Normal distribution with mean 5ms and variance 1ms, the fog to fog latencies are modelled by a Normal distribution with mean 10ms and variance 2ms, and device or fog to cloud latencies are modelled by another Normal distribution with mean 30 ms and variance 5ms. The containers modelling the devices, fogs, and cloud have equal weight in terms of the CPU scheduling.

\subsection{Parallel Results Push from Fogs}

The purpose of this experiment is to study the benefits of results sharing at the fogs. If the workload can be split among the fogs, we should see performance scaling up (or turn around times decreasing) as the number of fogs is increased.
For this experiment, we make use of the Wall-Following Robot Navigation Data dataset\footnote{https://archive.ics.uci.edu/ml/datasets/Wall-Following+Robot+Navigation+Data} from the University of California to perform machine learning using a feed-forward neural network. The time we measure in this experiment is the average of the time it takes a device to log data to the fog, perform machine learning at the fog and offload the trained network to a secondary application running on the same fog using the horizontal data sharing primitives built into the language. We run this experiments using a fixed number of devices (24 devices) and different number of fogs in order to evaluate the application load-balancing deployment feature built into the runtime which associates devices to fogs.

\begin{figure}[!ht]
\centering
\includegraphics[width=2.7in]{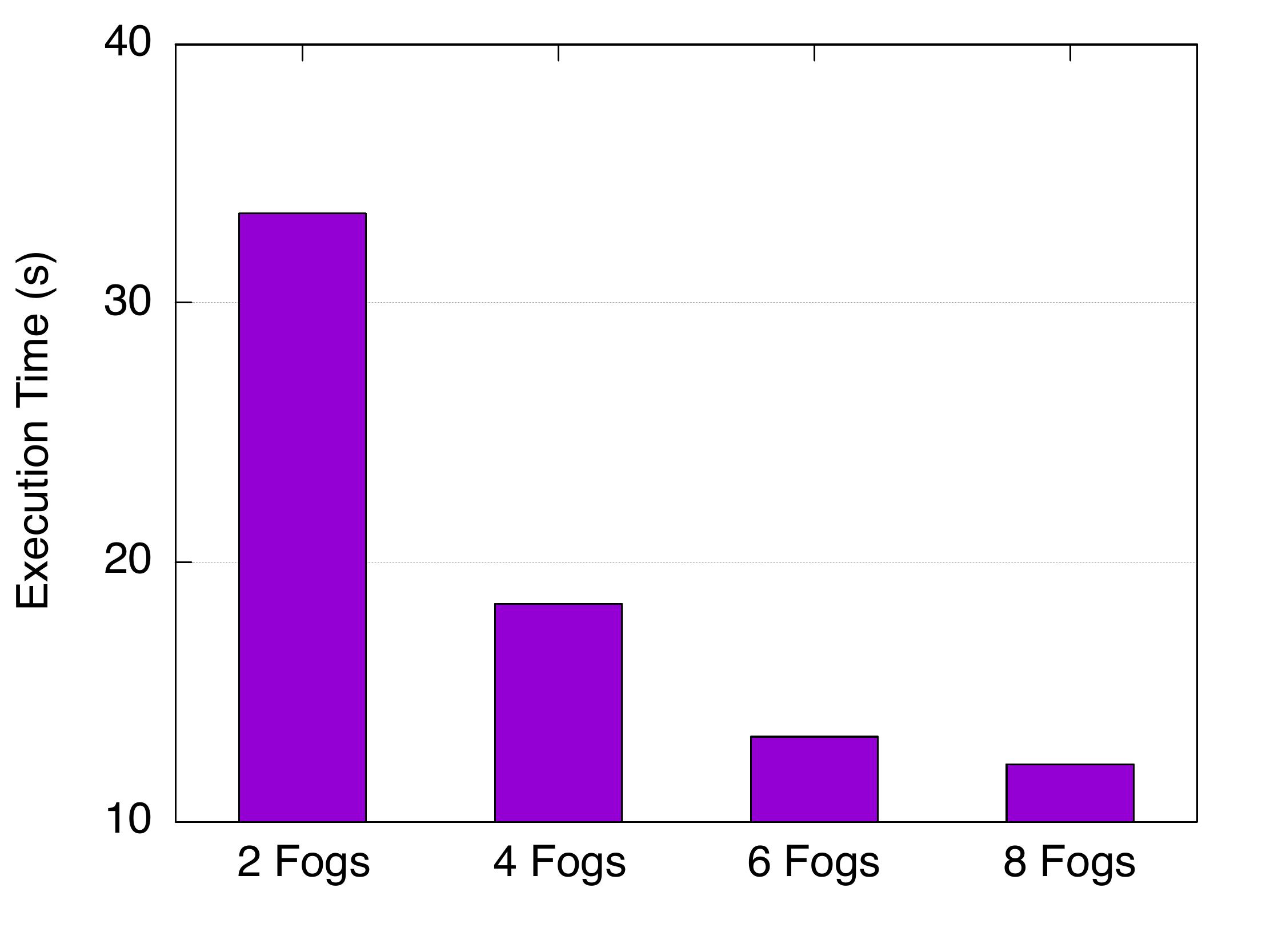}
\caption{Average time for data to move from a device through a feed-forward neural network on a fog to a secondary application.}
\label{measure1}
\end{figure}

In these experiments, the runtime was using a random fog assigning routine for load-balancing.
Figure~\ref{measure1} shows the result of this experiment. We observe that when we double the number of fog device, the execution time reduces by more than half. With the distributed memory model of the language, several applications can benefit from data generated in other applications. This language design feature enables large applications like the smart parking to be decomposed into several components.

\subsection{Turnaround Times and Cloud/Fog Levels}

\begin{figure}[!ht]
\centering
\includegraphics[width=2.9in]{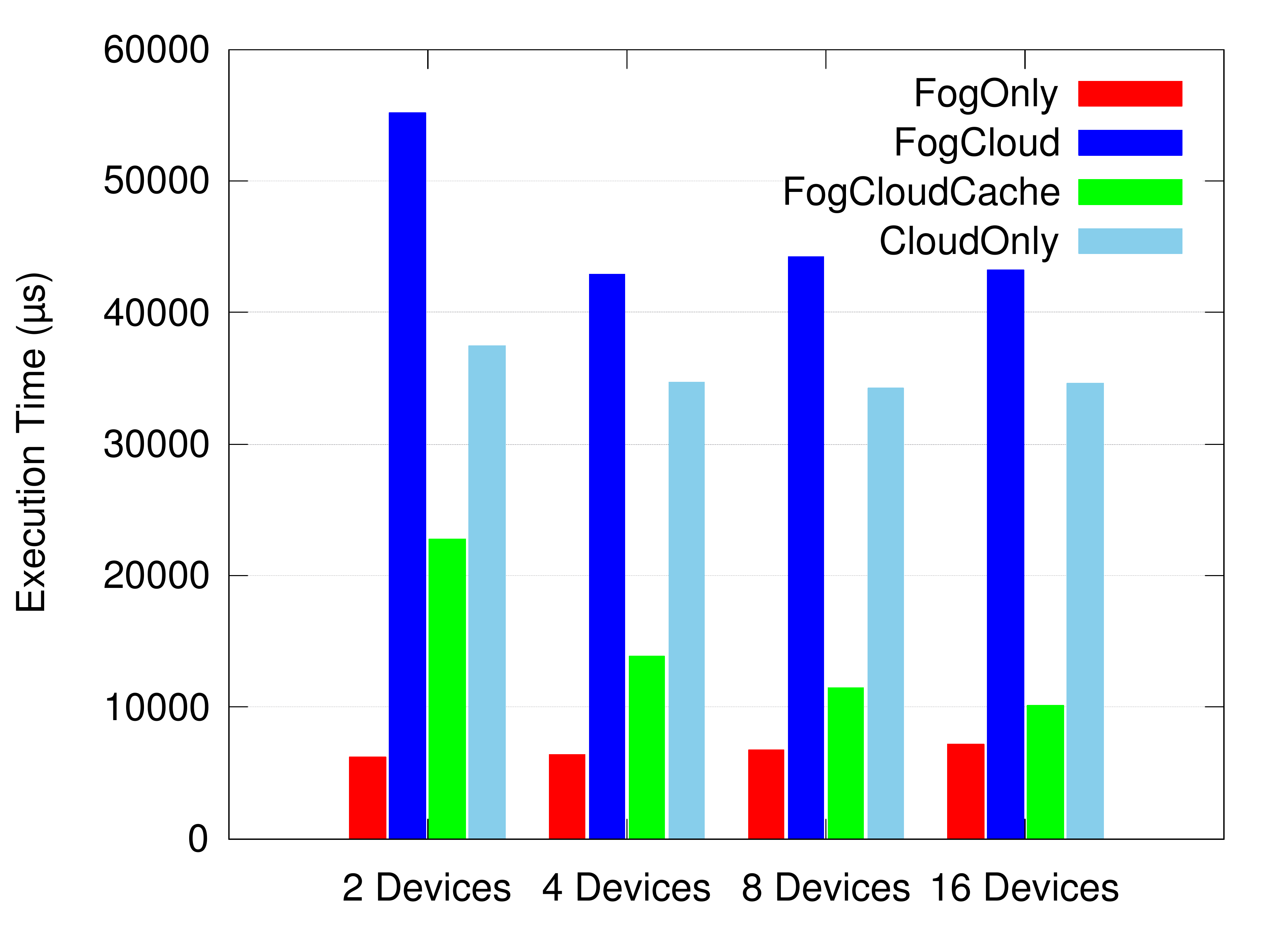}
\caption{Turnaround time for different data request scenarios using a single fog and cloud instance.}
\label{measure2}
\end{figure}

In this experiment, we want to demonstrate how tasks implemented in our language are able to obtain better performance from fogs for tiny tasks
compared to what could be obtained from the clouds.
As previously stated, our tree-based controller-worker programming model enables applications to be written once and deployed at different levels of the tree. In this experiment, we leverage the logger and broadcaster features of the language to determine the benefits of utilizing closeby data resources using turnaround time measurements. The result is shown in Figure~\ref{measure2}.

We create different scenarios to demonstrate this. First, we simulate a case where the data resources are available at the fog level - \textit{FogOnly}. In the second case which we refer to as \textit{FogCloud}, we consider a situation where an initial request is sent to the fog but the request is always routed to the cloud where the resource is located before the data is forwarded back through the fog to the requesting device.

Next, we present a modified version of the second case called \textit{FogCloudCache} that caches data obtained from the cloud on the fog and serves the cached data on subsequent requests. Finally, we examine a traditional device to cloud data resource request which we define as \textit{CloudOnly}. In each scenario, a single fog and cloud instance is configured for use with different number of devices. The execution time is the average round trip time for a device to request a resource from a top-level controller and receive a response.

The results show that application response is faster when resources are available at the fog level. With caching at the fog level, we see that the \textit{FogCloud} resource request method can be highly improved. The language runtime allows applications to benefit from closeby resources when available without the need to modify the code. Data requests are automatically sent to the immediate parent controller in the tree which can further request from a higher level when lower levels are unable to serve the request.

\subsection{Selective Logging of Data}
In order to demonstrate the benefits of the selective logging pattern built into the language, we generated a hypothetical dataset for the heart rate of a possible medical patient. We assume an application that detects abnormal heart rates above 100 beats per minute. It is important that such applications provide real-time data to interested parties to enable timely actions to be taken.

\begin{figure}[!ht]
\centering
\includegraphics[width=2.7in]{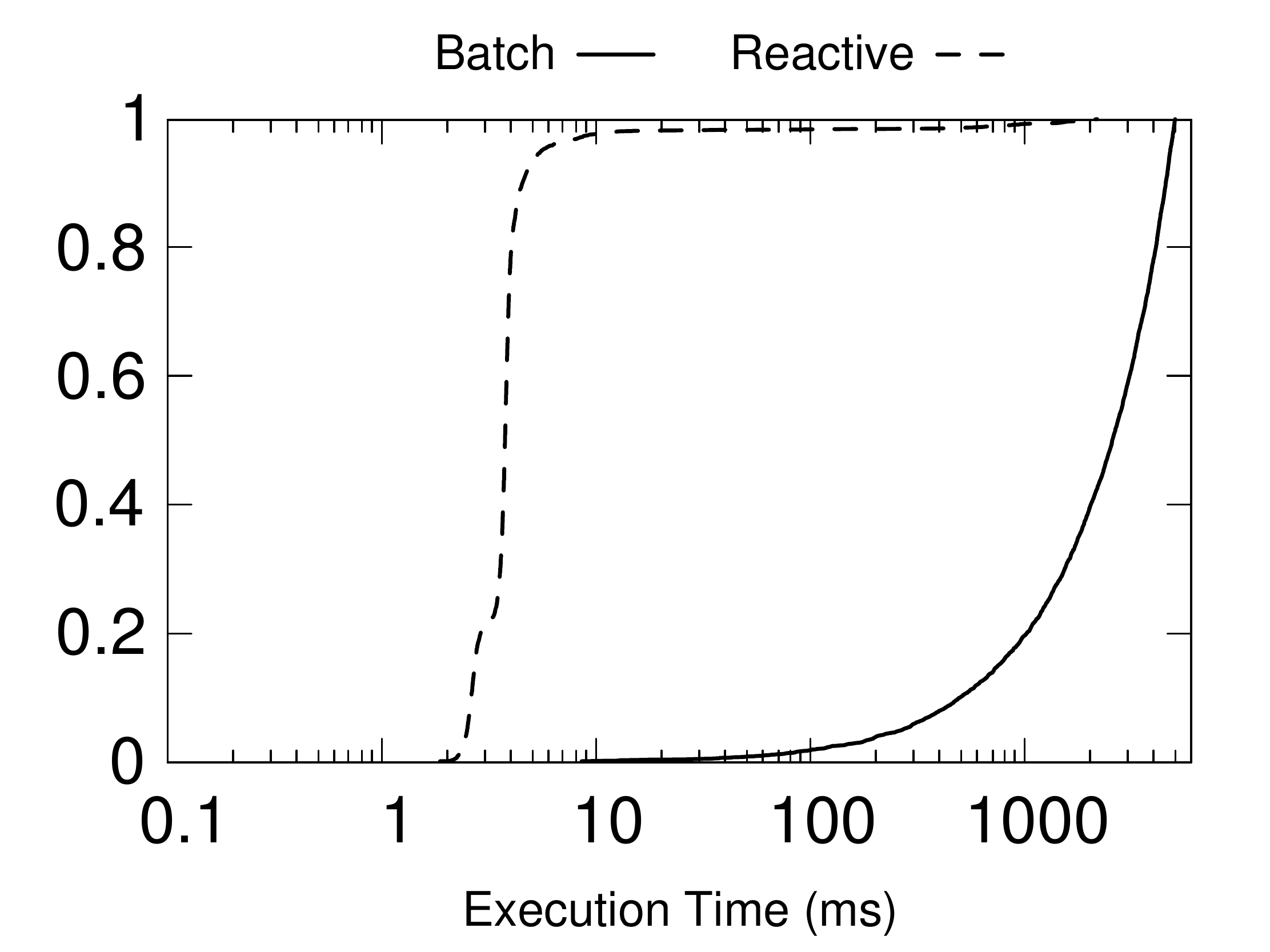}
\caption{Comparison between reactive processing with selective logging and batch processing.}
\label{measure3}
\end{figure}

We compare two types of applications: (i) One that uses the selective logging reactive mechanism to filter data in real time and forwards it to interested applications using the horizontal data sharing primitives and (ii) One the employs batch processing every 5 seconds to detect the abnormal data from queued logs before forwarding the detected abnormalities to listeners. The execution time is the time from data generation on the device level to the detection time on the fog. The CDF plot in Figure~\ref{measure3} shows that the selective logging with real-time filtering provides faster detection with more than 94\% of the detection falling around 5ms.

%% file: patterns.tex
\section{Edge-Oriented Programming Patterns}
\label{patterns}

In this section, we describe some interesting patterns that could be implemented in our language leveraging its unique edge-oriented design. The list of patterns provided here are not by any means exhaustive. An exhaustive enumeration of edge-oriented programming patterns would definitely help programmers harness this emerging resource stack.

\subsection{Function Placement Pattern}

When a controller calls a sub-controller, we could execute the call at any of the possible sub levels. For example, a call from the cloud could be executed at the fog or device. Similarly, when a worker calls a controller, we could execute that call in the device, fog, or cloud. The function placement pattern determines the exact location where the call actually executes. The function placement can be either static or dynamic. With static placement, we know at compile time where the function is going to run. We have either explicitly specified the location to run the function like device, fog, or cloud or we have left it to the default policy. The other option is to determine the location using conditions expressed on data values to determine the execution location.

In the example below, we use a {\tt jcondition} rule to specify that function {\tt deviceOnly} should run on a device. This particular definition is redundant because the default policy does the same execution, but it illustrates the use of explicit specification.

\begin{lstlisting}[language=JavaScript, caption=Sample \emph{jcondition} code fragment.]
jcond {
   	deviceOnly: sys.type == "device";
}
jasync function {deviceOnly} deviceOnly {
   	// Code to execute
}
\end{lstlisting}

We can use {\tt logger} and {\tt broadcaster} to express conditionals based on which a function execution location can
be decided.
Suppose we want to run a function at the edge (on the fog), but if the fog it too overloaded we want to fallback on the cloud.
We could implement such a condition as follows. We let the logger values push available load on the devices or fogs
and the broadcaster sends down the load threshold set
by the cloud.
\begin{center}
\begin{lstlisting}[language=JavaScript, caption=The \emph{loadCheck} and \emph{cloudRun} conditional rules.]
jcond {
   	loadCheck: sys.type == "fog" && load < 50;
   	cloudRun: sys.type == "cloud";
}
jdata {
   	int load as logger;
}
jasync function {loadCheck||cloudRun} loadBalanced {
   	// Code to execute
}
\end{lstlisting}
\end{center}

In the above code example, we use two {\tt jcondition} rules. The first rule \textit{loadCheck} will only run on a fog and uses the logger variable \textit{load} to check if the current load is below a fixed threshold. The load variable can be periodically updated elsewhere in the program by writing to the logger. The second {\tt jcondition}
rule allows the function to be run on the cloud. We use a logical or with the two rules when defining the function, ensuring the function will always run. Because functions will always try to run on the lowest possible level of the tree, if the loadCheck condition passes the function will run on the fog, and if it fails then the function will run on the cloud.

\subsection{Data Filtering Pattern}

Data filtering is a pattern for reducing the network bandwidth usage by transforming the data closer to the edge: at the devices or fogs. Our language provides a logger construct to push the data into the network. Data filtering can be applied quite easily by rewriting the data streams carried by the logger at the device or fogs using
application specific functions provided by the programmer.

For instance, if a vehicle (device) is equipped with multiple ambient temperature sensors it is not necessary to record all of them to get an estimate of the environmental temperature.
Suppose each temperature sensor is working as an independent C node with a corresponding logger instance,
the device J could compare the results of all sensors and perform a decision about which temperature reading is most likely to be correct. In our example, we will pick the one with the lowest temperature, as this sensor is likely to be least affected by sunlight. We can then send this data to the fog, which can perform a second filter. As each fog represents a local physical area it would not be necessary to relay the temperature readings from each device connected to it to the cloud for the same space. An average of all the reported temperatures from the devices can be taken to account for different calibrations in the sensors.
\begin{center}
\begin{lstlisting}[language=JavaScript, caption=Controller-side data filtering code snippet.]
jdata {
   	int tempSensors as logger(device);
   	int vehicleTemps as logger(fog);
   	int areaAverage as logger(cloud);
}
jcond {
   	deviceOnly: sys.type == "device";
   	fogOnly: sys.type == "fog";
}
jasync function {deviceOnly} vehicleTempFinder() {
   	var connectedDevices = tempSensors.size();
   	var lowestValue = tempSensors[0].getLastValue();
   	for(var i = 0; i < connectedDevices; i++) {
      		var sensorTemp =
        		connectedDevices[i].getLastValue();
        		if(sensorTemp < lowestValue) {
          		lowestValue = sensorTemp;
        		}
   	}
   	vehicleTemps = lowestValue;
}
jasync function {fogOnly} areaAverageCalculate() {
   	var connectedVehicles = vehicleTemps.size();
   	var sum = 0;
   	for(var i = 0; i < connectedVehicles; i++) {
      		sum += vehicleTemps[i].getLastValue();
   	}
   	areaAverage = sum / connectedVehicles;
}
\end{lstlisting}
\end{center}

The above code fragment shows the J side for data filtering.
On the C side, not shown, we would have each node writing the temperature values to the device logger \emph{tempSensors}. The \emph{vehicleTempFinder} function will run on the device J node because of the {\tt deviceOnly} rule. This function goes through the last logged state of the {\tt tempSensors} logger and finds the one with the lowest value. It logs the result to a fog logger \emph{vehicleTemps}, which stores the found value.
A fog only function \emph{areaAverageCalculate} takes the average of all the vehicle temperatures that are connected to that fog and saves it to the cloud logger \emph{areaAverage}.

Another form of data filtering is using the technique of {\tt selective logging}.
Instead of nodes' parents deciding which data is important from all the data that is sent,
we could select what data we would like to log.
In this case, we decide at a parent level such as the cloud, what criteria we are looking for in a device and tell it to turn on additional logging. For example, we could have vehicular devices that are logging their current position using GPS coordinates. The cloud would be connected to a live database of weather anomalies, such as smog. The cloud could then tell any vehicle that is within the GPS coordinates of a weather anomaly to turn on complete engine sensor logging so that a database of how the anomaly can affect engine performance can be created.

\subsection{Fog Fail-Over Pattern}\label{fogfail}

The Fog Fail-Over pattern deals with minimizing the disruption to the application from failures at the edge.
For example, if a device was using the fog to perform long calculations you would not want to restart the calculations from the beginning. We could handle this by periodically saving the progress of the calculations in a logger from the fog. If the fog would then crash, the device would automatically connect to a new fog, but all progress would be lost. The cloud could detect that progress has restarted based on the data being saved to the logger, and could push out the last saved update from the failed fog to the newly connected fog using the broadcaster. This pattern is partially
automated by the {\em shadow vertices} concept discussed in Section~\ref{patterns}.
\begin{center}
\begin{lstlisting}[language=JavaScript, caption=Code fragment showing simplified version fog fail-over pattern.]
jdata {
   	int saveState as logger;
   	int recoveryState as broadcaster;
}
jcond {
   	fogRun: sys.type == "fog";
   	cloudRun: sys.type == "cloud";
}
var computation;
jasync function {fogRun} fogCompute() {
   	// assign progress update

   	saveState.log(computation);

   	var oldProgress = recoveryState.getLastValue();
   	if(oldProgress > computation) {
      		computation = oldProgress;
   	}
}
var saveStateLog;
jasync function {cloudRun} cloudCheck() {
   	var newUpdate = saveState[0].getLastValue();
   	if(newUpdate) < saveStateLog) {
      		recoveryState.broadcast(saveStateLog);
   	} else {
      		saveStateLog = newUpdate;
   	}
}
\end{lstlisting}
\end{center}

In the above code sample, we show a simplified version of the example described above. We assume there is only one device, one fog and one cloud. In the sample code (for illustration), we use an {\tt int} type logger to save
progress. The \emph{fogCompute} function performs a calculation that is not shown, and saves the result to the cloud logger \emph{saveState}. The cloud runs a function \emph{cloudCheck} that checks if the state of the logger is further along than the previous result. If the result is less than the previous result, then the cloud can assume that the fog has failed and a new fog was brought in, restarting the computation. It then broadcasts the last recorded value back to the fog. The fog checks each time it writes to the log if the cloud has broadcasted a new greater value, showing that there was a previous fog further along in computation, and replaces its progress with the broadcasted value. When we add more devices and fogs to this example, we would have to include an identifier code in the logger so that the cloud would know which fog it is comparing progress with.

\subsection{Pipes and Filters at the Edge Pattern}

The Pipes and Filters at the Edge pattern is an adoption of the well known {\em Pipes and Filters} pattern.
In our pattern, each J node (cloud, fog, or device) could concurrently communicate with the partner component
to interchange data. This way the amount of data that can be interchanged can increase if the applications at the pipe endpoints have more overlapping nodes. Typically, the interchanges happen at the fog level because the devices may not be overlapping. The fogs are the most likely overlapping nodes between the two partner applications that are interconnected by the pipe.
Taking advantage of this pattern, we can separate functionality into different sub-applications.
Another major advantage of this pattern, the {\em disconnection tolerance}.

For example, we could have an application that runs on all the smart lights in a room and a separate application that runs on smartphones. When the user connects the phones to the fog that the lighting is on, the applications could communicate with each other at the fog level using the pipes and filters pattern. When the user changes the state of the lighting on their phone, the message would be sent up to the fog using a logger. The fog would then use a in/out flow (key mechanism implementing the interchange necessary for the pipes and filters) to send a message to the application controlling the state of the lights, which would use a broadcaster to send a message to all lights in the room to change their state.

%% file: applications.tex
\section{Sample Applications: Design \& Experiences}
\label{samples}

In this section, we describe two applications that were designed and implemented using our language. The applications were tested on a single machine and in a docker-based emulator for a cloud, fog, device system.

\subsection{Distributed Shell}

By running multiple instances of a program written in our language, we can quite easily create a distributed system. Those instances self-organize into a cloud-fog-device hierarchy and connect with each other assuming there is connectivity among the resources running the different instances. With mobiles at the edge, we can expect a collection of volatile and heterogeneous nodes that can be at different states of engagement with the global system. The goal of the Distributed Shell (referred to as DShell) is to create a command-line interface (CLI) backed by our language runtime to tackle the management of a volatile set of nodes. It enhances the programmability of the language by reducing the bootstrapping effort needed while deploying the applications written in the language. It allows parallel (on a tree structure) mass deployment of programs to nodes.

Consider a drone fleet that needs management. We want DShell to run on all of them and the edge nodes (fog stations) and cloud.  Through DShell, it is possible to deploy patches and execute subprograms in parallel on connected nodes. The organization of the fleet in a tree enables an intuitive traversal of the clouds, fogs, and devices in the network is made possible by the shell, as well as specific targeting of individual nodes and subtrees.

The DShell design leverages the non-blocking execution capabilities of the language.
This is important because our language is single-threaded; the shell should not halt the collective network awaiting a response from a given node. In DShell, this requirement
is realized through the use of {\tt jasync} non-blocking functions in the language, in which requests are fired off towards connected nodes without blocking the calling node. Correspondingly, request responses can be received through {\tt jcallback} functions allowing a node to perform C2J and J2C asynchronous communications.

Once instantiated and fully connected, the DShell is a tree rooted at the cloud. When it experiences disconnections, there can be many components of the DShell each not connected to the other. However, as the network disconnections heal and the nodes are able to restore connectivity to each other, the DShell will automatically restore itself to the connected state and become a tree that is rooted at the cloud again. To accomplish this self-healing, DShell relies completely on the self-healing nature of the underlying runtime, which uses a global and local discovery service to detect possible nodes to reconnect at the different levels. At least in its current design there is no additional processing beyond the default runtime processing performed by the language.

Like a UNIX shell, one of the key tasks of DShell is to start, stop, and keep track of jobs, which are the execution of other programs written in our language. Any arbitrary program written in our language could be instantiated as a ``tree,'' which mirrors the structure of the DShell itself. However, the size and shape of the tree structure of an executing program depend on the number of nodes that are part of the execution. For instance, if only one node is executing a program, the tree contains only a single J node and possibly a single C node. When DShell is asked to execute another program, it spawns a child of the calling DShell process. Unlike the case of a UNIX shell where spawning creates a single child, in DShell, the spawning could create a subtree of processes. The subtree of processes so spawned will be responsible for loading and running the application DShell was asked to execute.

There are several advantages for using DShell to launch a program that is written in our language instead of using the native shell. A native run would create new instances of all resources such as Redis store and MQTT brokers that are needed for the execution of the program. Whereas DShell would reuse the resources already allocated to the parent shell and assign the new application a separate namespace so that the new application is isolated from the parent shell and also from other programs that may already be running within the shell. One of the restrictions of running a program from within DShell is that the new program needs to launch on a subset of the resources or whole of the resources that are already held by the parent DShell.

\begin{figure}[!ht]
\centering
\includegraphics[width=2.5in]{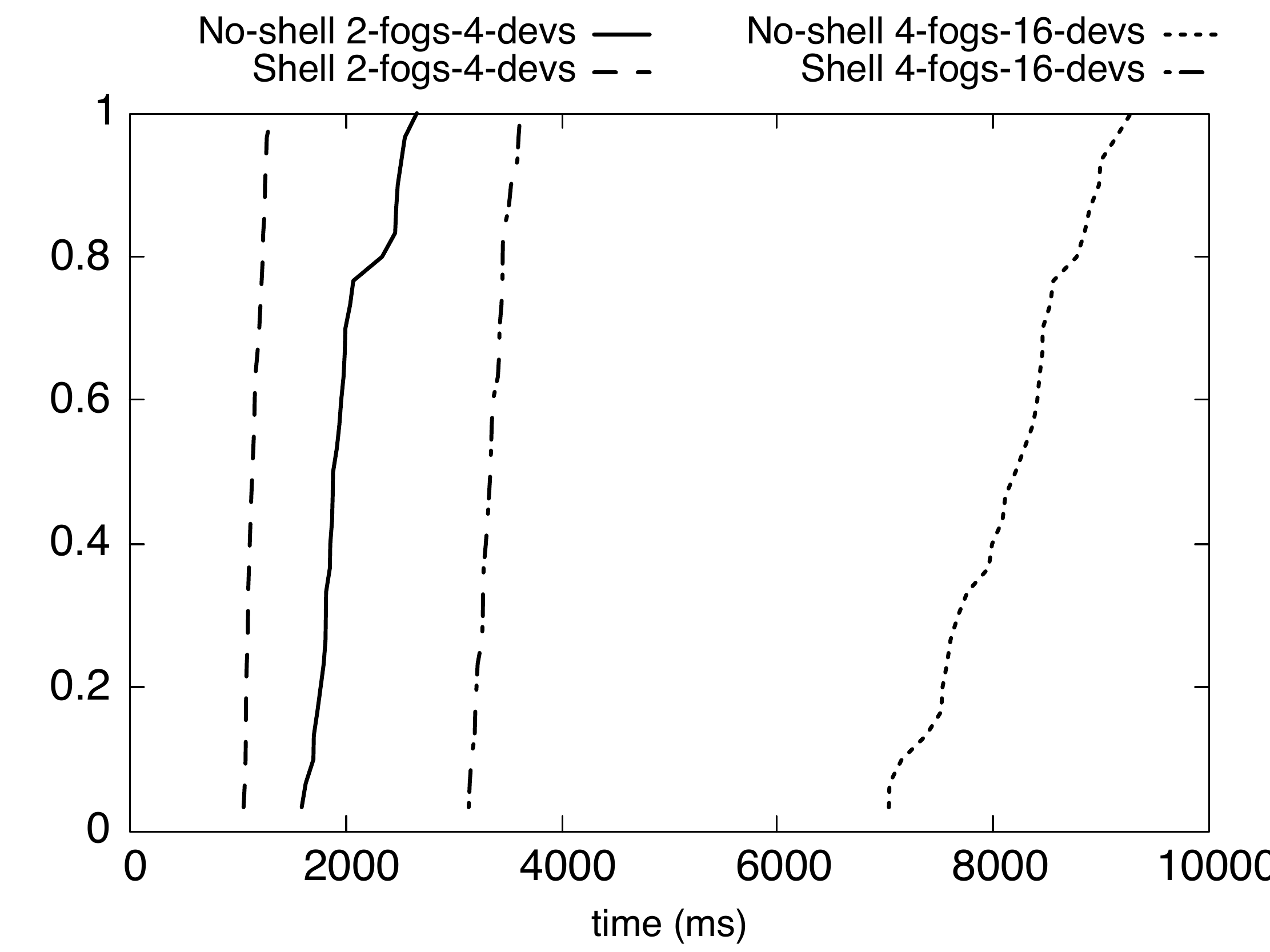}
\caption{CDF showing the startup time for two different tree configurations with shell and without shell.}
\label{shell}
\end{figure}

Figure~\ref{shell} shows the startup times of launching programs from inside the shell and natively on different configurations of nodes. These experiments were all performed on a single host with 4 cores. A future evaluation will use the docker-based emulator.

The results indicate that DShell is much quicker in launching applications. The main reasons for the reduced startup times are: reuse of key resources to execute the application and also parallel spawning in different subtrees.

With volatile nodes (i.e., nodes that change state often), it is quite
challenging to maintain state in DShell regarding the job execution state across the distributed shell. In our language, persistent node-to-node data flows are implemented through broadcasters and loggers, respectively allowing top down and bottom up data transfer.

Another feature implemented in DShell is the Pipes and Filters at the Edge pattern described in Section~\ref{patterns}. Using this feature we can start an application and pipe its output to another program that is spawned to run on the same resource pool. Using this pattern, we can decompose a problem into smaller sub problems and create solutions individually and then compose a complete solution by running the sub programs in a pipe according to the data flow requirements.

\subsection{Parking Spot Manager}

\begin{figure}[!ht]
\centering{\includegraphics[width=3.3in, height=2.1in]{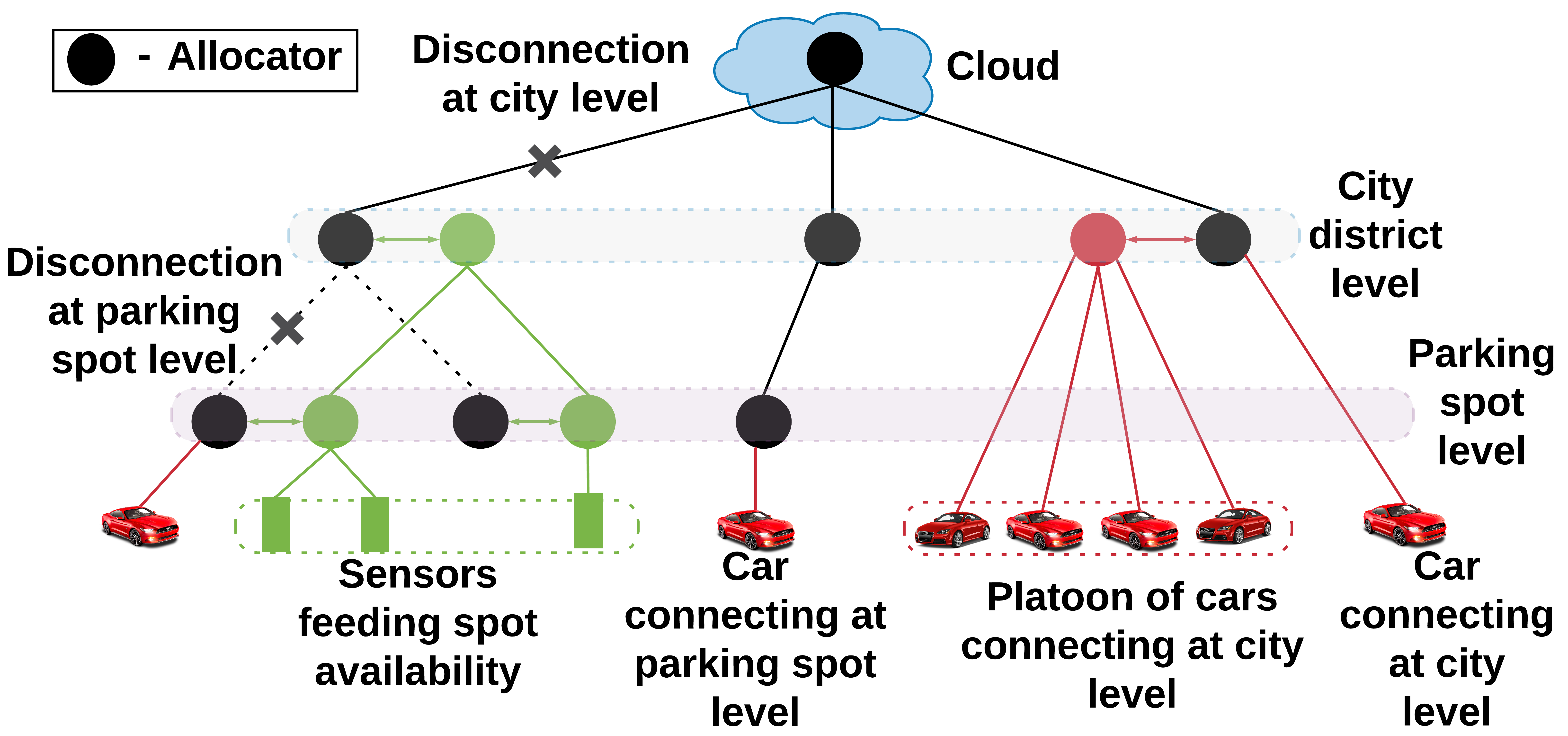}}
\caption{Overall architecture of the parking spot manager.}
\label{parking}
\end{figure}

Our language allows programmers to implement data-intensive applications in different ways by enabling
a wide range of edge-oriented data flow possibilities that can provide higher QoS in terms of response times
for requests. In order to demonstrate these features, we implemented a parking spot manager application which receives requests for premium parking services and responds with information about available spots within the requested area and makes suggestion for other available spots in nearby areas when none exists within the preferred location. We opted for a parking spot application for two reasons:
\begin{enumerate}
 \item It is non-trivial and is inherently distributed, has time-constrained processing requirements and is
 representative of many applications in smart city scenarios.
\item  The implementation cuts across many features of the language discussed previously.
\end{enumerate}
The parking spot manager application is a stack of applications that consists of several smaller applications that exist at the device, fog and cloud levels. Figure~\ref{parking} shows the constituent applications and the data flows among them.

The different components of the parking spot manager are:
\begin{enumerate}
 \item {\em Car}: The car application is made up of two logical parts: (i) one that models the execution in the device (car in this case) and (ii) component that runs in the fog. The first part consists of a C-Node module which receives location instructions about an available spot sent from the fog or cloud that can be routed to the car's navigation system, a J-Node module which communicates with the C-Node, as well as the car's dashboard where the request can be triggered from by either using a voice command or touch pad. We model a touch pad dashboard with a simple web application that communicates with the J-Node module using socket.io. As previously mentioned, applications in our language can be deployed at different levels without modification which makes the second part technically the same as the first. The second part consists of the same J-Node application module being executed at the device level; however, the conditional constructs allow execution restrictions for user-defined functions to be enabled at either the device, fog or both levels and thus makes it runnable at the fog.
\item {\em Sensing}: This component models the detection and sensing of cars on parking spots and informs the fog component of the application about the occupancy status and properties of each spot within its coverage. The J-Node application at the device level can be responsible for 1 or more spots within a parking lot. Ideally, such an application could place a sensor at each spot which is modelled as a C-Node at the device level.
\item {\em Allocating}: This is the application that ties all the different application components together and is responsible for receiving requests and forwarding allocations to requesting cars. This application exist at both the fog and cloud levels.
\end{enumerate}

%% file: related.tex
\section{Related Work}
\label{related}

\subsection{Software Defined Programming}
Software defined networking has inspired a host of software defined programming frameworks for IoT \cite{jararweh2015sdiot, liu2015software, tortonesi2016spf}, edge \cite{jararweh2016sdmec}, fog \cite{truong2015software, gebre2017mist} and cloud computing \cite{nastic2015rtgovops}. A software defined IoT framework encompassing software defined storage \cite{wu2013software} and software defined security \cite{al2015sdsecurity} was developed in \cite{jararweh2015sdiot}. The framework's model includes the following layers, the physical layer consisting of the physical devices in the system, the middleware layer where the IoT and software defined controllers reside and the application layer. A similar three-layered architecture was proposed in \cite{liu2015software} for smart urban sensing. 

MIST \cite{gebre2017mist}, a software defined fog system leveraging software defined mobility \cite{contreras2016software} for real-time surveillance was developed to provide device mobility and change adaptively based on context. Nodes are classified as being controllers, hosts or switches to enable seamless execution even in the presence of frequent changes in membership. A runtime framework for managing software defined IoT clouds was proposed in \cite{nastic2015rtgovops}. The goal of the framework is to abstract dynamicity and scalability issues from the management of software defined IoT cloud systems. The framework provides an automated and fine-grained central control and an autonomous IoT cloud resources aimed at improving system efficiency.  

\subsection{Cloud Programming}
Orleans \cite{bykov2011orleans} is a software framework for building scalable and reliable cloud applications. Orleans uses an actor-based distributed components called grains that consist of isolated units. Grains communicate asynchronously through message passing and use a single-threaded execution model. In cases of high load and lower system throughput, Orleans creates multiple instances of a busy grain to handle the busy grains multiple simultaneous executions. The consistency model in Orleans is achieved using optimistic and atomic transactions that are isolated. One of the main design goals of Orleans is that the runtime handles system level attributes such as scalability, reliability, fault tolerance and durability while application developers only focus on the application logic.

Dripcast \cite{nakagawa2014dripcast} is a Java based application development framework to integrate smart devices into cloud computing infrastructure. Further, it is a server-less framework for storing and processing Java objects in a cloud environment. These Java objects will be made available on smart things and users can manipulate those objects as if they are local objects. It implements transparent Java remote procedure call and a mechanism to read, store, and process Java objects in a distributed, scalable data store.

\subsection{IoT Programming}

Mobile Fog \cite{hong2013mobile} is a platform as a service programming framework for IoT. An interface with a single code is exposed that allows dynamic scaling. A Mobile Fog application consists of processes that cover defined geographical locations. Processes are connected in a three-tier hierarchical architecture and are distributed on computing elements such as cloud, fog and edge devices. Assumptions made in Mobile Fog are that fog computing infrastructure nodes are placed in the network and a programming interface is provided by the fog. Mobile Fog handles scaling by making application developers specify the scaling policy at each hierarchical level. Load balancing is based on creating on-demand fog instances at the same level as an over-loaded fog instance.

Calvin \cite{persson2015calvin} is a framework that merges IoT and cloud in a unified programming model. It is an IoT programming framework, which combines the ideas of actor model and flow based computing. To simplify application development, it proposes four phases to be followed in a sequential fashion: describe, connect, deploy, and manage. These phases are supported by the runtime, APIs and communication protocols. The platform dependent part of Calvin runtime manages inter-runtime communication, transport layer support, abstraction for I/O and sensing mechanisms. The platform-independent runtime provides interface for the actors. The scheduler of the Calvin runtime resides in this layer. Calvin runtime supports multi-tenancy. Once an application is deployed, actors may share runtime with actors from other applications.

Simurgh \cite{khodadadi2015simurgh} provides a high level-programming framework for IoT application development. The framework supports exposing of IoT services as RESTful APIs and
composing the IoT services to create various flow patterns in a simplified manner. The overall Simurgh architecture has two main layers: things layer and platform layer. In the ``thing layer'' are network discovery and registration broker that listens to incoming connection requests from devices` and handles them. The ``platform layer'' stores information about things and services, manages flow design and composition, and handles requests. An API mediator assists programmers to expose their applications through RESTful APIs. The Simurgh framework provides detailed support to IoT development. Assistance to develop, manage and reuse flow patterns as provided in this framework is crucial for IoT programmers.

\subsection{Mobile Computing Programming}

Odessa \cite{ra2011odessa} proposes a mobile programming model where application developers structure applications as data flow graphs. Odessa is best suited for streaming applications where automatic data offloading and parallelism are desired attributes. An edge or connector in a data flow graph represents data dependencies between the vertices that represent different stages of processing. Processing stages in an application operate independently with no shared states and only communicate through connectors thus abstracting the programming details and complexity of different processing stages from one another. Odessa's programming design is based on the Sprout runtime \cite{pillai2009slipstream} which allows developers to write parallel and distributed applications while hiding the complexity behind them.

\subsection{Comparisons with our Programming Language and Middleware}

Inspired by SDN, our language and middleware follow a controller-worker model with multiple levels of controllers. Unlike software defined programming frameworks where the software defined controller has full autonomy over orchestrating the activities of workers, our programming language and middleware requires only a single program to be written for both the controllers and workers. The program is split at compile time into the controller and worker components. A bi-directional relationship exists between controllers and workers. Our language and middleware not only provide a programming interface like existing programming frameworks in IoT and cloud, but includes an allocation optimization model for careful assignment of workers to controllers.

Similar to Orleans \cite{bykov2011orleans}, our programming language is single threaded and allows multi-application tenancy as in Calvin \cite{persson2015calvin}. Our language provides different programming patterns for dealing with issues such as fault tolerance, reliability and data manipulation unlike frameworks such as \cite{bykov2011orleans} where optimistic and atomic transactions are used to provide reliable and fault tolerant executions. Our language is not restrictive to certain types of applications like in \cite{ra2011odessa} which is most suited for streaming applications. Our language provides support for writing different types of applications ranging from data-intensive applications to control and actuation based applications.

%% file: conclusions.tex
\section{Conclusions and Future Work}

To enjoy the full benefits of edge computing, we need applications that are aware of the edge computing setup and adapt their processing accordingly. The first step in creating such applications is the development of programming languages and patterns that considers edge computing as a key element. This paper does just that. It presents a new polyglot programming language for a cloud-fog-device system. We considered many problems that are paramount in
edge computing while designing and implementing the language such as: function placement at the appropriate levels to achieve desired QoS, on-demand data probing to reduce the bandwidth usage, fault tolerance at the fog and load distribution at the edge. 

Our language is based on a distributed-node programming model. To handle the data-intensive nature of IoT related computing tasks, a flow-based data management scheme is built into the language to handle up flow and down flow of data in the tree. One of the challenges of engaging the fogs is fault tolerance. In this paper, we present an fog-to-device allocation optimizer that works with our language design. Essentially, this optimizer determines the best fog-to-device mapping by finding an optimal partition of the data graph created by the logger objects representing the devices. The optimizer design presented in this paper is not yet part of the language implementation. As part of the future work, we will implement it. All necessary software hooks for implementing the optimizer like request router is already implemented. We also describe several edge-oriented programming patterns that can leverage our language design. In this paper, we present some patterns that relate to the key design concerns. 

We conducted three experiments as part of evaluating the performance of the language and its runtime: 
(i) machine learning based experiment performing device-to-fog data logging, (ii) experiment to evaluate the benefits of selecting logging that implements data filtering at the edge versus cloud, 
(iii) experiment to evaluate the latencies of using resources at different levels.
To further demonstrate the applicability and functionality of our language, we developed two applications. The first is a distributed shell application created for easily running, stopping, and managing multiple instances of a program written in our language and the second is a parking spot manager consisting of several smaller applications written in our language for advertising parking spot availability to cars.

As part of the future work, we will continue to improve the runtime of the language by integrating efficient synchronization facilities. The existing synchronization capabilities in the language for controller-to-workers is pretty basic. We are also planning on writing and testing more applications in our language and deploying some of the applications in a real-world scenario.